\documentclass[prb,aps,twocolumn]{revtex4}
\usepackage{amsfonts}
\usepackage{amsmath}
\usepackage{amssymb}
\usepackage{graphicx}

\begin{document}

\title{First Principle Noncollinear Transport Calculation and Interfacial Spin-flipping of Cu/Co Multilayers}

\author{Ling Tang and Shuai Wang}

\affiliation{State Key Laboratory for Surface Physics, Institute of
Physics, Chinese Academy of Sciences, P.~O.~Box 603, Beijing 100080,
P.~R.~China.}

\begin{abstract}
In this paper the first principle noncollinear transport calculation
for Cu/Co(111) including interfacial spin-flipping was performed. We
modeled spin-flipping at the interface by assuming a noncollinear
magnetic structure with random magnetization orientation which
satisfied Gaussian distribution along average magnetization
direction. The relationship between spin-dependent conductance
including interfacial spin-flipping and random magnetization
orientation distribution width was obtained. For certain
distribution width, our defined spin-flipping ratio coincides with
the range of experimental spin-flipping probability
$P=1-e^{-\delta}$, where $\delta=0.25\pm0.1$. The magnetoresistance
in Co/Cu/Co spin valve system including interfacial spin-flipping
has also been calculated.
\end{abstract}

\maketitle

\section{Introduction}

The electron transport across ferromagnetic/nonmagnetic (FM/NM)
interface is of interesting in the past two
decades.\cite{history1,history2,history3} Based on the two current
model(non spin-flipping theory), $ab$ $initio$ calculation with
no-free parameter of interfacial specific resistance\cite{xia-prb01}
agreed reasonably well with experimental data for some
lattice-matched metal pairs, such as fcc
Cu(111)/Co(111)\cite{bass-mmm99a} and bcc
Fe(110)/Cr(110).\cite{bass-mmm02}

The spin-flipping at the FM/NM interface, which certainly exists in
the experiment, is not well studied, partially due to the
theoretical difficult, and partially due to lacking of reliable
experimental data.\cite{bass-jpc07} However, spin-flipping at the
FM/NM interface is getting increasing importance. Spin-flipping at
the ferromagnetic/supperconductor interface can induce spin triplet
pairing in the ferromagnetic
side.\cite{xing-prl07,Efetov-prl03,lofwander-prl05} In addition, the
spin-flipping at the FM/NM interface can also change the spin
torques induced by current.\cite{Levy-prb06,MD-06,slon}

Moreover, Geux $et$ $al.$ have calculated the transmission
probability in the presence of magnetic impurity scattering with
spin-flipping by the effective mass approximation.\cite{bauer} They
calculated the transmission probability matrix in spin space and
obtained that the spin-flipping probability is proportional to the
impurity density. They also found that to the first order the
calculated conductances from transmission probability matrix
decrease linearly with increasing the impurity density. However, the
transport properties through the real interface with interfacial
spin-flipping process have not been studied yet.

In this paper, we will calculate the scattering matrix of real FM/NM
interface with spin-flipping process by the first principle
noncollinear transport calculation.\cite{shuai} The spin-flipping at
the interface is modeled by assuming a noncollinear magnetic
structure with random magnetization orientation which satisfied
Gaussian distribution along average magnetization direction. By the
noncollinear transport calculation,\cite{shuai} we can obtain how
the interfacial conductance changes with random interfacial magnetic
structure and the effect of interfacial spin-flipping on the
magnetoresistance in FM/NM/FM spin valve system.

\section{Computational Details}

Our calculation of scattering matrix is based on the surface Green's
function method\cite{ies} with tight-binding linear muffin tin
orbital basis.\cite{mto} First, the self-consistent one-electron
effective potential in our calculation is obtained from collinear
electron structure calculation\cite{ies} without spin-orbit
coupling. The Hamiltonian $\hat{H}_{0}$ is constructed by this
self-consistent potential and is diagonal in spin space. Second, the
rigid potential approximation has been used in our noncollinear
transport calculation. In this approximation, we rotate the
Hamiltonian $\hat{H}_{0}$ which is in local quantum axis
representation in spin space to the global quantum axis
representation. So the Hamiltonian for transport calculation can be
written as
$\hat{H}'=\hat{U}(\theta,\varphi)\hat{H}_{0}\hat{U}^{\dag}(\theta,\varphi)$,
where $\hat{U}(\theta,\varphi)$ is the unity rotation matrix in spin
space. $\theta$ and $\varphi$ are the polar angle and azimuth angle
of the local quantum axis respectively (global quantum axis is taken
as z axis and $\varphi=0$ in our calculations). Therefore the
spin-flipping is only induced by interfacial magnetic disorder and
the spin-orbit coupling is neglected in our transport calculation.

\begin{figure*}
  \includegraphics[width=17.2cm]{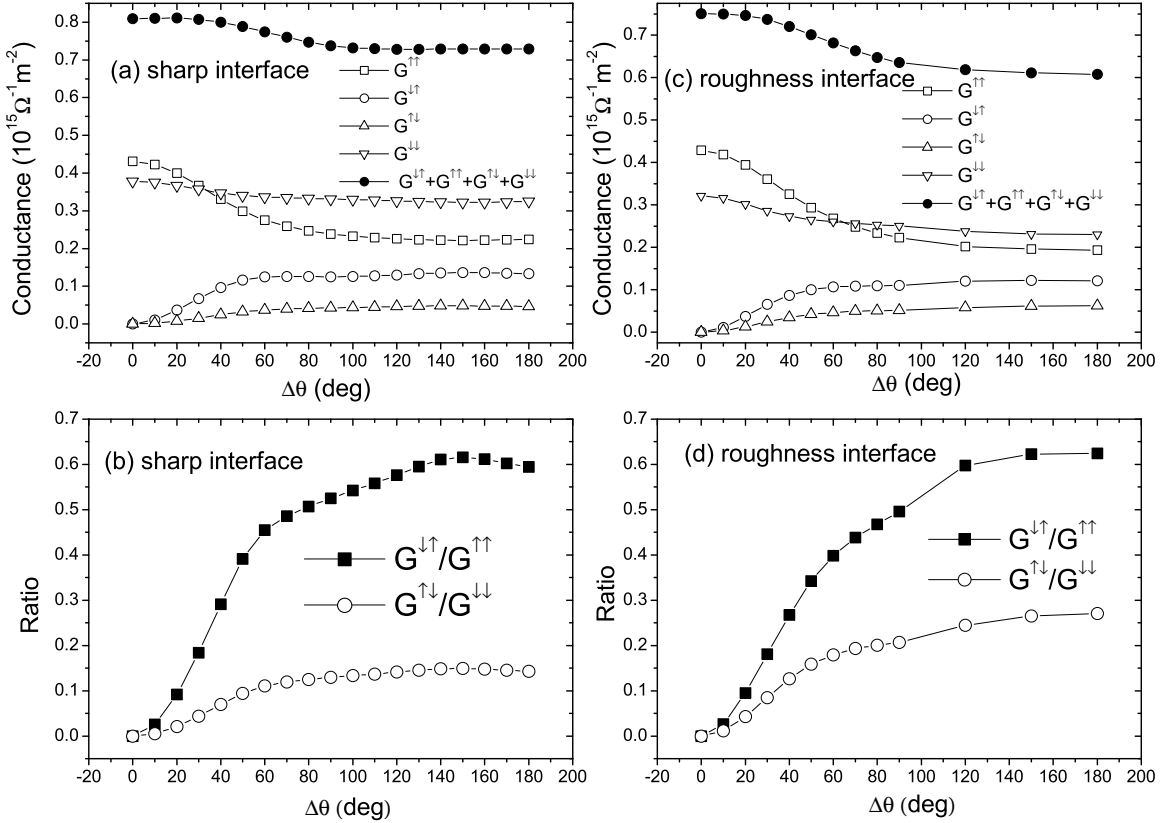}\\
  \caption{The spin-dependent conductances of sharp and roughness
Cu/Co(111) interfaces. (a) and (c) The conductances will be
saturated for large $\Delta\theta$ case and the critical
$\Delta\theta$ of roughness interface is larger than that of sharp
interface. (b) and (d) For both sharp and roughness interface the
spin in majority channel can be flipped more easily than that in
minority channel.}\label{f1}
\end{figure*}

Due to the lack of the details of interfacial magnetic disorder, we
assume that spin-flipping is introduced by the Gaussian random
distribution of magnetization orientation. In our calculation all
the magnetization orientation of Co atoms are along one global
quantum axis except the Co monolayer at the interface. Further the
disordered magnetization orientation is modeled by $10\times10$
lateral supercell. The deviation angles within the $10\times10$
supercell satisfy the Gaussian random distribution, where the
average orientation is along global quantum axis and the
distribution width is $\Delta\theta$. So the most deviation angles
in the supercell are in the range of (-$\Delta\theta$,
$\Delta\theta$). However, the magnitude of magnetization for each Co
atom in the supercell is constant. In this paper we have calculated
the transport properties for different magnetization orientation
distribution width, which is from $\Delta\theta$ = 0 to 180 (deg).
Here the larger distribution width $\Delta\theta$ implies the larger
probability of spin-flipping scattering process at interface.

The noncollinear magnetic structure at interface has been calculated
by other group using first principle local spin density
calculations,\cite{Oparin-jap99} and the multiple and metastable
noncollinear magnetic structure have been obtained in
copper-permalloy interface. There are four metastable noncollinear
states in which the energies are lower than the energy of collinear
state or total random state. The collinearity of those four
metastable states in their paper\cite{Oparin-jap99} is in the range
of about 0.4 to 0.8, which corresponds to our parameter
$\Delta\theta\approx$ 40 to 70 (deg).

\section{Results and Discussion}

In our noncollinear Cu/Co(111) transport calculation the lattice
constant is taken as $a$=3.549{\AA}. The spin-dependent conductance
is
\begin{equation}\label{eq1}
    G^{\sigma\sigma'}=\frac{e^{2}}{h}\sum_{\mu, \nu, \mathbf{k}_{\|}}
    T^{\sigma\sigma'}_{\mu\nu}(\mathbf{k_{\|}})=\frac{e^{2}}{h}\sum_{\mu, \nu, \mathbf{k}_{\|}}|t_{\mu\nu}^{\sigma\sigma'}(\mathbf{k_{\|}})|^{2}
\end{equation}
where $t_{\mu\nu}^{\sigma\sigma'}$ is transmission matrix element
for bloch state ($\nu, \sigma'$) in lead Cu to bloch state ($\mu,
\sigma$) in lead Co and $\mathbf{k}_{\|}$ is lateral wave vector.
Figure.\ref{f1} shows the spin-dependent conductances with different
magnetization orientation distribution width $\Delta\theta$, where
the roughness interface is modeled by 2ML of 50\%-50\% alloy in a
10$\times$10 lateral supercell, which can be denoted as
Cu[Cu$_{0.5}$Co$_{0.5}$$|$Cu$_{0.5}$Co$_{0.5}$]Co. For this
roughness interface the random Gaussian distribution of
magnetization direction only takes place at the interfacial magnetic
atoms (Co atoms). Here $G^{\uparrow\uparrow}$ and
$G^{\downarrow\downarrow}$ are conductances for majority and
minority electron channel with unchanging the orientation of spin.
$G^{\downarrow\uparrow}$($G^{\uparrow\downarrow}$) is conductance
for spin-flipping process, which describes the probability of
majority (minority) being scattered to minority (majority).

As shown in Figure.\ref{f1}(a) and (c), for both sharp and roughness
interfaces, with increasing the magnetization direction distribution
width $\Delta\theta$ the spin-flipping conductances increase and the
majority (minority) conductance decreases, where the majority
conductance decreases more rapidly and the conductance
$G^{\downarrow\uparrow}$ increases also more rapidly than
$G^{\uparrow\downarrow}$. For sharp interface with increasing
$\Delta\theta$, the total conductance
($G^{\downarrow\uparrow}+G^{\uparrow\uparrow}+G^{\uparrow\downarrow}+G^{\downarrow\downarrow}$)
increases slightly at first and decreases to constant when
$\Delta\theta>90$ (deg). But for the roughness interface, the total
conductance decreases monotonically and the critical distribution
width $\Delta\theta$ where the conductance start to be saturated is
about 120 (deg), which is larger than that for sharp interface. In
our results of electron structure calculation, the average
interfacial magnetic moment per atomic sphere is about 0.77$\mu_{B}$
for roughness interface and 1.58 $\mu_{B}$ for sharp interface.
Moreover for the same distribution of random interfacial magnetic
structure, the larger interfacial magnetic moment will lead to more
strongly spin-flipping scattering of the incoming electron.
Therefore, it needs much more degree of interfacial magnetic
structure disorder to saturate the conductance in the roughness
interface.

\begin{figure}
  \includegraphics[width=8.6cm]{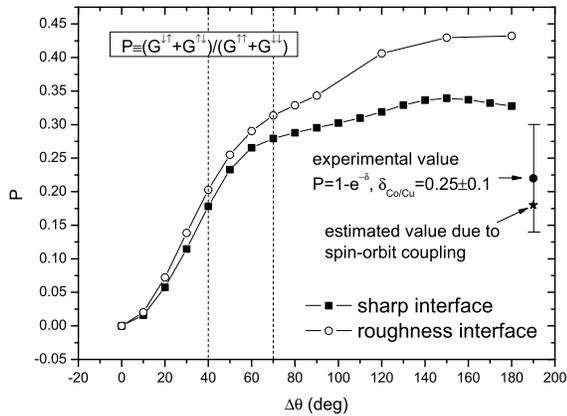}\\
  \caption{The spin-flipping ratio across Cu/Co(111). The solid circle
is the spin-flipping probability value
$P=1-e^{-\delta}\approx1-e^{-0.25}\approx0.22$ inferred from
experiment data and the error bar comes from the uncertainty
$\delta=0.25\pm0.1$. The solid star is the estimated spin-flipping
probability only due to the spin-orbit coupling. The range of
$\Delta\theta$ between dash line corresponds to the collinearity 0.4
to 0.8 which is result of the magnetic structure
calculation.}\label{f2}
\end{figure}

In the case of non spin-flipping Cu/Co(111) interface the minority
electron is reflected more strongly than the electron in majority
channel. Taking into account the scattering induced by noncollinear
magnetization, the conductances $G^{\uparrow\uparrow}$ and
$G^{\downarrow\downarrow}$ both decrease with increasing
$\Delta\theta$. Further the influence of noncollinear magnetization
scattering is relatively more important for majority channel than
for minority channel. Therefore as shown in Figure.\ref{f1}(a) and
(c), $G^{\uparrow\uparrow}$ decreases more rapidly than
$G^{\downarrow\downarrow}$. In addition, for both sharp and
roughness interfaces we observe that
$G^{\downarrow\uparrow}/G^{\uparrow\uparrow}>G^{\uparrow\downarrow}/G^{\downarrow\downarrow}$
in Figure.\ref{f1}(b) and (d), which also indicates that the spin in
majority channel can be flipped more easily than that in minority
channel.

\begin{figure}
  \includegraphics[width=8.6cm]{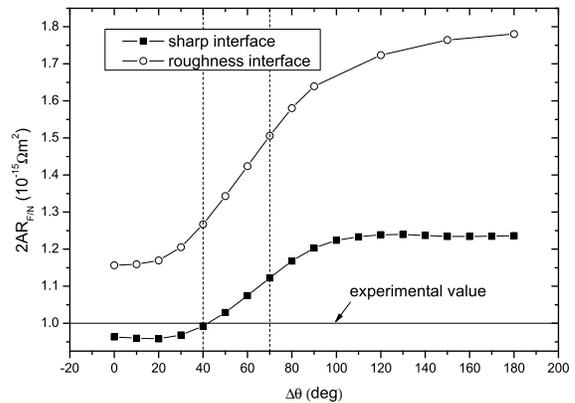}\\
  \caption{The rounded interfacial specific resistance $2AR_{F/N}$ for
different distribution width $\Delta\theta$, where the line in the
figure corresponds to the value from the experiment. The region
between two dash line corresponds to the result of magnetic
structure calculation.}\label{f3}
\end{figure}

The interfacial spin-flipping parameter $\delta$ usually describes
the interfacial spin-memory-loss which is defined as
$\delta=t_{I}/l_{sf}^{I}$ within VF theory,\cite{fert} and the
spin-flipping probability is\cite{bass-jpc07} $P=1-e^{-\delta}$. For
Cu/Co interface $\delta\approx0.25$ which is indirectly inferred by
explaining the difference between 'interleaved' and 'separated'
sample within VF theory.\cite{bass-prb02,bass-mmm01,bass-jap03b}
There is no direct measurement of parameter $\delta$ for F/N
interface up to date and the value of $\delta$ inferred from
experiment is quite uncertain.\cite{bass-jpc07} In this paper, we
defined the spin-flipping ratio as
$P\equiv(G^{\downarrow\uparrow}+G^{\uparrow\downarrow})/(G^{\uparrow\uparrow}+G^{\downarrow\downarrow})$.
From Figure.\ref{f2} one can see that for $\Delta\theta\approx$ 40
to 70 (deg) our defined spin-flipping ratio coincides with the range
of spin-flipping probability inferred from experimental data
($P=1-e^{-\delta}$, $\delta=0.25\pm0.1$).\cite{bass-prb02} This
might be just an accident coincidence because we have not calculated
the spin-flipping probability directly and our definition of the
spin flipping ratio is only used for giving some quantitative
information of spin-flipping. In addition, K. Eid $et$
$al.$\cite{bass-prb02} estimated the interfacial spin-flipping
probability at Cu/Co interface only due to the spin-orbit coupling
and the result is $P=1-e^{-\delta}\approx1-e^{-0.2}\approx0.18$,
which is on the small side of the experimental $P$ value range. The
estimated spin-flipping probability due to spin-orbit coupling
approximates the P value for the case of $\Delta\theta=40$ (deg) in
our calculation and is much smaller than the saturated spin-flipping
probability due to magnetic disorder. Considering our calculation
only taking the interfacial magnetic disorder into account, it is
suggested that the spin mixing effect of interfacial magnetic
disorder is more prominent than that of spin-orbit coupling on the
Cu/Co transport properties, e.g. current-perpendicular-to-plane
(CPP) magnetoresistance (MR).

\begin{figure}
  \includegraphics[width=8.6cm]{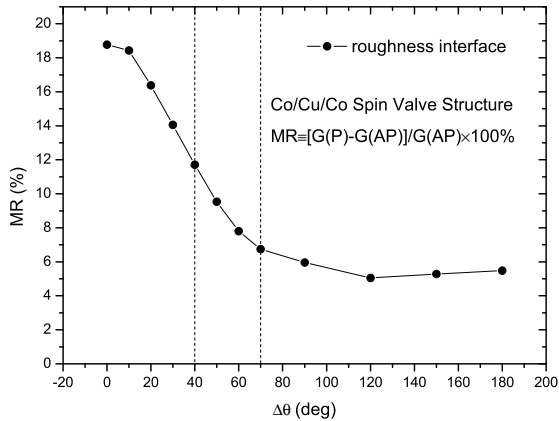}\\
  \caption{The calculated MR of the spin valve system Co/Cu/Co with
different interfacial spin-flipping, where the two Cu/Cu interfaces
are both roughness with 2ML 50\%-50\% alloy. The region between two
dash line also corresponds to the result of magnetic structure
calculation.}\label{f4}
\end{figure}

Figure.\ref{f3} shows the rounded interfacial specific resistance
which is defined as\cite{schep-97}
\begin{equation}\label{eq2}
    2AR_{F/N}=2A\frac{h}{e^{2}}[\frac{1}{\sum
    T^{\sigma\sigma'}_{\mu\nu}}-\frac{1}{2}(\frac{1}{2N_{\textmd{Cu}}}+\frac{1}{N_{\textmd{Co}}^{\uparrow}+N_{\textmd{Co}}^{\downarrow}})]
\end{equation}
where $A$ is section area, $N_{\textmd{Cu}}$,
$N_{\textmd{Co}}^{\uparrow}$ and $N_{\textmd{Co}}^{\downarrow}$ are
the Sharvin conductances. First, in non spin-flipping case for sharp
interface we obtain $2AR_{F/N}\approx0.96$ ($10^{-15}\Omega m^{2}$)
which is close to the experimental value\cite{bass-mmm99a} $\sim1.0$
($10^{-15}\Omega m^{2}$). Second, due to the additional scattering
from noncollinear magnetization, with increasing $\Delta\theta$ the
rounded interfacial specific resistance increases and for large
distribution width the calculated $2AR_{F/N}\approx1.23$
($10^{-15}\Omega m^{2}$) which is about 128\% of non spin-flipping
specific resistance. Moreover the experimental specific resistance
value corresponds to the case of $\Delta\theta\approx40$ (deg) in
our calculation, which indicates that the spin-flipping at interface
can also explain the deviation of rounded interfacial specific
resistance between sharp interface calculation and experimental
data. The rounded interfacial specific resistance of roughness
interface is also shown in Figure.\ref{f3}. It can be seen that the
specific resistance of roughness interface also increases with
increasing $\Delta\theta$ as the case of sharp interface. However,
the value of roughness interface is larger than that of sharp
interface and the $\Delta\theta$ in which the specific resistance
start to be saturated is also larger than that of sharp interface.

Figure.\ref{f4} shows the magnetoresistance(MR) of the spin valve
system Co/Cu/Co with roughness interface for different interfacial
spin-flipping. We have calculated the conductances of the parallel
G(P) and antiparallel G(AP) configuration, and the MR is defined as
$\textmd{MR}\equiv
\frac{\textmd{G(P)}-\textmd{G(AP)}}{\textmd{G(AP)}}\times100\%$. One
can observe that the MR decreases rapidly with increasing
distribution width $\Delta\theta$ and for $\Delta\theta>90$ (deg)
the MR is nearly constant about 5\%. For the region of
$\Delta\theta$ = 40 to 70 (deg) corresponding to the result of
magnetic structure calculation,\cite{Oparin-jap99} the MR is in
range of 11.7\% to 6.7\%, which is about 2/3 to 1/3 of the non
spin-flipping MR $\sim$ 18.8\%.

\section{Conclusion}

In this paper the first principle noncollinear transport calculation
for Cu/Co(111) including interfacial spin-flipping was performed. We
modeled spin-flipping at the interface by assuming a noncollinear
magnetic structure with random magnetization orientation which
satisfied Gaussian distribution along average magnetization
direction. The relationship between spin-dependent conductance
including spin-flipping conductance and random magnetization
orientation distribution width was obtained. We found that the
conductances start to be saturated when the distribution width
$\Delta\theta$ larger than the critical width. For distribution
width $\Delta\theta$=40 to 70 (deg) our defined spin-flipping ratio
coincides with the range of spin-flipping probability
$P=1-e^{-\delta}$, where $\delta=0.25\pm0.1$ inferred from
experimental data. In addition we also found that the
magnetoresistance of Co/Cu/Co spin valve system decreases rapidly
with increasing interfacial spin-flipping probability.

{\bf Acknowledgements} The authors acknowledge Prof. Ke Xia for
suggesting the problem and the help from Yuan Xu about the
calculation. We gratefully acknowledge financial support from
NSF(10634070), MOST(2006CB933000, 2006AA03Z402) of China.

\end{document}